\documentclass[11]{article}

\usepackage[top=0.5in,bottom=0.5in]{geometry}
\usepackage{graphicx}
\usepackage{amsfonts}
\usepackage{wrapfig}
\usepackage{subcaption}
\usepackage{xcolor}

\begin{document}
\title{\textbf{``Dark matter'' neutrinos: A proposed mechanism for the stellar rotational velocity curve of the Milky Way}}
\author{A.Panteli}
\date{\today}
\maketitle
\begin{center}
\begin{flushleft}
\textbf{Email:~andrewpanteli@tiscali.co.uk~~~~~~~~~~~~~~~~~~~~~\textsc{Prepared for submission to  JCAP}}
\end{flushleft}
\end{center}
\thispagestyle{empty}
\setcounter{page}{1}

\noindent \textbf{Abstract}. The inconsistency of the observed rotational velocity curve of the Milky Way with the current theory of gravitational dynamics is well documented. The curve implies the presence of significantly more mass, or ``dark matter'', than is observable. Current models estimate the amount of dark matter to be$\mathrm{\sim }$85\% of total ``matter''. This paper explores relativistic neutrinos as a potential source of dark matter and proposes a crude binary star accretion model mechanism at the centre of the Milky Way for their production. The results show this to be consistent with the current theory of gravitation, the observed velocity curve of the Milky Way and current estimates of the proportion of dark matter to baryonic matter$\mathrm{\sim }85\%$.\\

\noindent \textbf{Introduction}\\

\noindent It is well documented \cite{sofue2001rotation} that the observed rotational velocity curve for the Milky Way, as shown in Fig.1 for example, is not consistent with current gravitational theory applied to observed baryonic mass. For this to be the case, one would expect rotational velocities to reduce according to the inverse square root of their distance from a distance enclosing most of the baryonic mass of the galaxy, at $\sim 10kpc$ and beyond. However, these velocities appear to remain relatively rather uncorrelated with marginal increasing distance from the edge of the central bulge and beyond the region containing most of the detectable stellar and baryonic mass (Fig.1). This suggests the potential existence of a relatively dominant component of galactic mass that appears to be largely proportional to the distance from the centre rather than tending to a Keplerian curve. Farther out, beyond 20kpc, observed velocities appear to drift lower, tending to a "floor" value in the mid hundreds of kms$^{-1}$,rather than tailing off.\\

\noindent In order to account for the non-Keplerian observed velocities, various dynamic models have been constructed on the basis of fitting the masses of the three galactic components-- bulge, disk and a dark matter ``halo'', to observed velocities. This approach has resulted in generating significant implied, yet physically unspecified, dark matter halo masses of$\sim 85\%$ of total galactic mass \cite{sofue2012grand}. The aim of this paper is to propose a physical mechanism for the creation of a dark matter halo that is consistent with current understanding of the physical laws of gravitation and neutrinos.\\
\begin{figure}[h]
	\centering
	\begin{minipage}[b]{.48\textwidth}
		\includegraphics[width=0.98\linewidth]{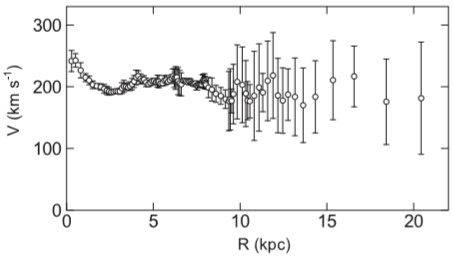}
		\caption{Observed curve up to 20kpc.\cite{sofue2001rotation}}	
	\end{minipage}
	\begin{minipage}[b]{.48\textwidth}
		\includegraphics[width=0.98\linewidth]{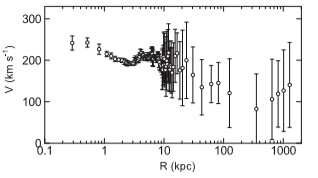}
		\caption{Observed curve up to 1000kpc.\cite{sofue2001rotation}}	
	\end{minipage}
\end{figure}

\noindent \textbf{The Model}\\

\noindent A relatively crude model was employed for the purpose of readily assessing whether the results would be of the order of magnitude required to satisfy the objective of the exercise, thereby warranting further investigation, rather than to consider second order effects and/or a more complex approach that would require resources beyond the scope of the exercise. The relatively unsophisticated approach taken notwithstanding, results appear to be consistent with current physical laws and knowledge.\\

\noindent It was assumed that the disk of the Milky Way contains \textit{N}${}_{B}$$\mathrm{\sim }$1.65x10${}^{10}$${}^{\ }$ stars and the central bulge of the Milky Way is composed of \textit{N}${}_{D}\mathrm{\sim}$3.41x10${}^{10}$ \cite{sofue2012grand} relatively older Population II stars \cite{baade1951galaxies} of mass \textit{M}${}_{\mathrm{\odot }}$ with an average main sequence lifespan of $10Gyr$ and, given the relatively dense environment and short fall-in times, that such bulge stars would be in binary systems with other stars initially of similar age and mass. It was assumed that an accretion disk structure would form about the primary component of the binary at the time of the secondary leaving the main sequence by way of Roche Lobe overflow due to the level of viscosity/turbulence of the region, with the primary and secondary being potentially interchangeable. The Eddington mass flow rate of consumption of the accretion disk of the primary, $\dot{M}{}_{p}$, is given by:

$$\dot{M}{}_{p}=\frac{4\pi GM{}_{p}m{}_{p}}{\mathrm{c}\mathrm{\sigma{}_{p}}},$$

\noindent where $\sigma{}_{p}$=proton cross-section of interaction$\mathrm{\sim }7x10{}^{-29}m{}^{2}; G\mathrm{\sim }6.7x10{}^{-11}Nm{}^{2}kg{}^{-2}; M{}_{p}$=mass of primary; $m{}_{p}$=mass of proton$\mathrm{\sim }2x10{}^{-27}kg$; $c\mathrm{\sim }3x10{}^{8}ms{}^{-1}$.\\

\noindent 

\noindent Given that fall-in times are much less than consumption times, the time taken for a primary to consume the mass of a secondary was taken as M/$\dot{M}\mathrm{\sim }0.4Gyr$ and that after such time it would consume the secondary, doubling its mass after allowing for the mass radiated during the accretion process and subsequently enter the next Phase. This process would then repeat with a frequency once per $\mathrm{\sim }0.4Gyr$ until the current Phase, Phase IX, at $t=13.5Gyr$, which is estimated to have started at $t=13.3Gyr$. M${}_{p}$ was initialised to 1\textit{M}${}_{\mathrm{\odot }}$ at $t=10Gyr$. The proportion, $\eta$, of the accreted mass radiated during each Phase, is given by:

$$\eta = \frac{GM_p}{c^2R},$$  

\noindent  for $R=5R_s$, where $R_s=$ Schwarzschild radius. The mass of a primary at the start of Phase Z, $M^z_p$, given the mass of a primary at the start of Phase I of \textit{M}${}_{\mathrm{\odot }}{}_{}$, is given by:
$$M^z_p{} = M{}_\odot(2-\eta)^{(z-1)},$$

\noindent and the temperature of the accreting mass is given by \cite{padmanabhan2002theoretical}:\\

\noindent $$T{}_{a}=7.3x10{}^{7}(M{}_{p}/{M}{}_\odot){}^{-0.25}$$

\noindent It can be seen from above that the effective temperatures associated with the accretion matter range from $\sim 73x10^6K$ for a Phase I primary of mass $\sim 1M_\odot$ to $\sim 20x10^6K$ for a Phase IX primary of mass $170{M}{}_\odot$ -- exceeding the critical temperature required for nucleosynthesis.\\ 

\noindent Assuming a nucleosynthetic process, whereby a proportion, $\psi$, of radiated mass $\sim 0.02$ takes the form of ``solar'' neutrinos of energy $E^\nu_\odot\sim530keV$~\cite{ianni2014solar} produced at a core solar temperature $T_\odot\sim 15x10^6K$, the energy of neutrinos radiated from a Phase IX accretion disk is given by $E^\nu_\odot = T_{a}/T_\odot\sim714keV$. Assuming a rest mass of $\sim 0.0132eV/c^2$~\cite{palanque2015neutrino} for electron neutrinos in their normal hierarchical state, this would result in the production of relativistic neutrinos with a Lorentz gamma of $\gamma \sim 6.4x10^7$. The number of primaries in Phase I at time $t=10Gyr$ is initialised at $N^1_p \sim 2.48x10^{10}$ to generate a corresponding baryonic bulge mass at $t=13.5Gyr$ of $1.65x10^{10}M_\odot$\cite{sofue2012grand}. From the above it can be seen that the neutrino mass flux in Phase Z, $\Phi^z_\nu$, in $kgyr^{-1}$, is given by:

$$\Phi^z_\nu=\frac{4 \pi G m_p \eta \psi \gamma N^1_p M{}_\odot(2-\eta)^{(z-1)}}{c\sigma_p}3x10^7$$
\noindent The corresponding neutrino mass enclosed within a sphere of radius $r$ lyrs from the galactic centre, $M^z_\nu$, is simply given by:

\noindent $$M^z_\nu = \Phi^z_\nu r$$

\noindent The raw neutrino mass model above is used to compute the mass function of the neutrino sphere, or halo, for consecutive increasing values of $r$. It can be seen that this function is linear in $r$, which is a key condition for a flat velocity curve.\\

\noindent The mass function of the raw neutrino model was scaled over a bulge distance corresponding to 0$\mathrm{<}$$r$$\mathrm{<}$7kly to take into account the baryonic mass function of the bulge, given that neutrino production would be directly proportional to the baryonic mass of the binaries within the bulge. The bulge mass function M${}_{B}$(r) developed to fit the bulge mass of the above for a radius of $r$=7kly from the galactic centre is: \\

\noindent $$M_B(r) = \sum^{7000}_{r=1} \frac{\rho_0}{r^{2.43508}}V_rM_\odot,$$ \\
for $0<r<7kly$, where $\rho_0=10^7pc^{-3}$ at $r=1pc$~\cite{ryden2003} and $V_r$ is the volume of the $r^{th}$ shell of the sphere. The mass function of the disk M${}_{D}$(r) is assumed to be linear from $r_B$=7kl to $r_D$=30kly, enclosing the mass given by the above, and is simply given by:
\noindent $$M_D(r) =  \frac{M_D}{(r_D-r_B)}r,$$ 

\noindent where $r_B$, $r_D$ are the radii of the bulge and disk respectively. It can be seen from the above, and after scaling the bulge neutrinos, the total mass function M${}_{T}$(r), for Phase IX with Z=9, is given by the following four terms:
$$M_T(r) = \sum^{7000}_{r=1} \frac{\rho_0}{r^{2.43508}}V_rM_\odot+$$

$$\sum^{7000}_{r=1}\frac{\rho_0}{r^{2.43508}}V_r\frac{M_\odot}{M_B}\frac{4 \pi Gm_p\eta \psi \gamma N^1_pM_\odot(2-\eta)^8}{c{\sigma }_p}3x10^7r+$$

$$\frac{M_D}{(r_D-r_B)}r+$$ 
 
$$\frac{4 \pi Gm_p\eta \psi \gamma N^1_pM_\odot(2-\eta)^8}{c{\sigma }_p}3x10^7r$$

\noindent corresponding to the baryonic bulge mass and neutrino bulge mass for $0kly<r<7kly$, baryonic disk mass for $7kly<r<30kly$ and neutrino non-bulge mass for $7kly<r<129kly$ respectively.\\

\noindent \textbf{Results}\\
\noindent The above total mass function M${}_{T}$(r) is used to compute the rotational velocities $v$, assuming bound orbits, simply from:
$$v=\sqrt{\frac{GM_T(r)}{r}}$$
\begin{figure}[h]
	\centering
	\begin{minipage}[b]{.45\textwidth}
		\includegraphics[width=0.98\linewidth]{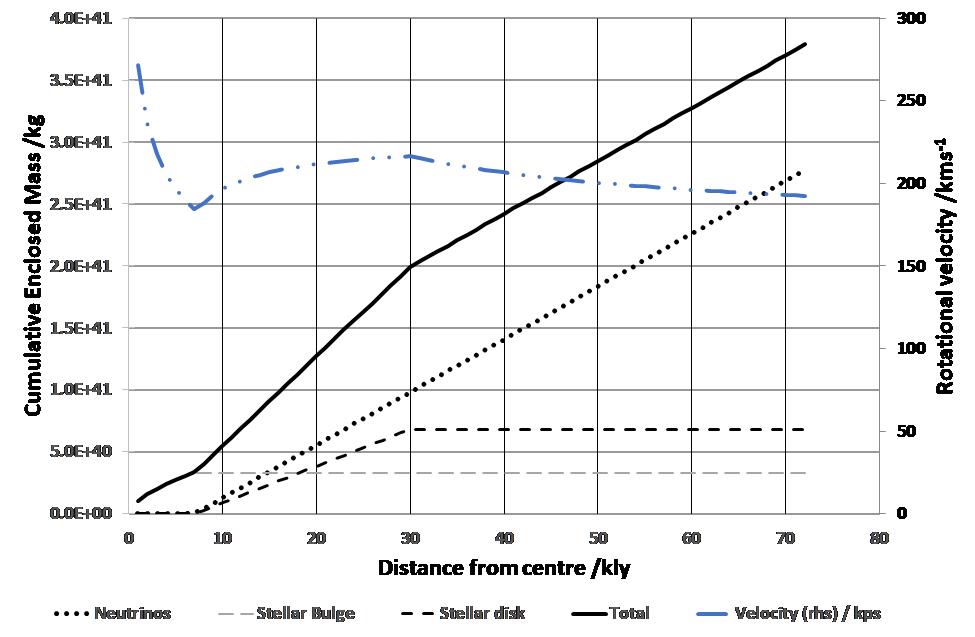}
		\caption{Mass-velocity profile 70kly.}	
	\end{minipage}
	\begin{minipage}[b]{.48\textwidth}
		\includegraphics[width=0.98\linewidth]{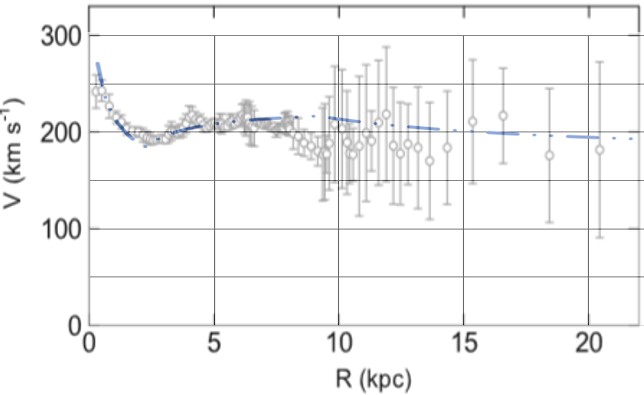}
		\caption{Est. v. observed velocities 20kpc.}	
	\end{minipage}
\end{figure}

\noindent The resulting estimated velocity curve shown in Fig.3 is compared with the observed velocity curve in Fig.1 for r=0 to$\sim70kly$. The two are shown superimposed in Fig.4.\\
\noindent Similarly, Fig.5 and Fig.6 show the corresponding mass-velocity functions over a distance of one galactic radius, assumed to be $\sim129kly$, and the superimposition of the resulting velocity curve on the observed velocities respectively.

\begin{figure}[h]
	\centering
	\begin{minipage}[b]{.45\textwidth}
		\includegraphics[width=0.98\linewidth]{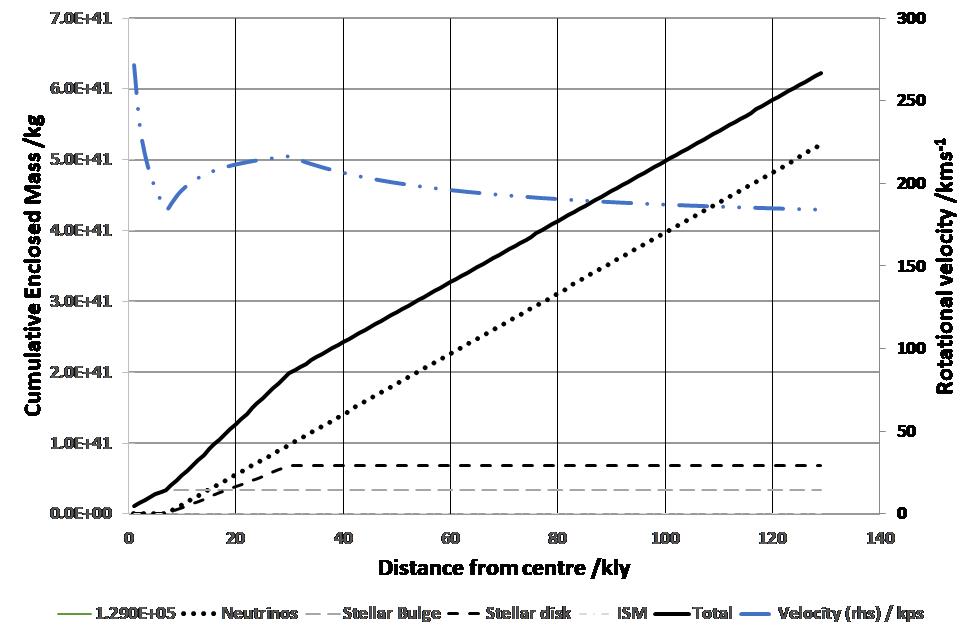}
		\caption{ Milky Way: Mass-velocity profile.}	
	\end{minipage}
	\begin{minipage}[b]{.49\textwidth}
		\includegraphics[width=0.98\linewidth]{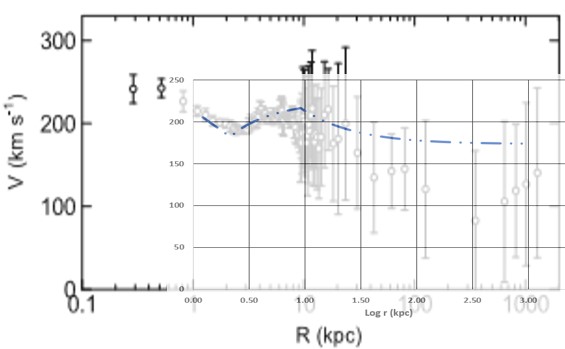}
		\caption{Est. v. observed velocities 1Mpc.}	
	\end{minipage}
\end{figure}

\noindent A table of the resulting estimated masses and velocities as a function of radius are shown in Fig.8 and a summary of the model mechanism is shown in Fig.9. A breakdown of the proportion of the total enclosed estimated mass by bulge, disk and neutrino sphere is shown in Fig.7.\\

\begin{figure}[h]
	\centering
	\begin{minipage}[b]{.46\textwidth}
		\includegraphics[width=0.99\linewidth]{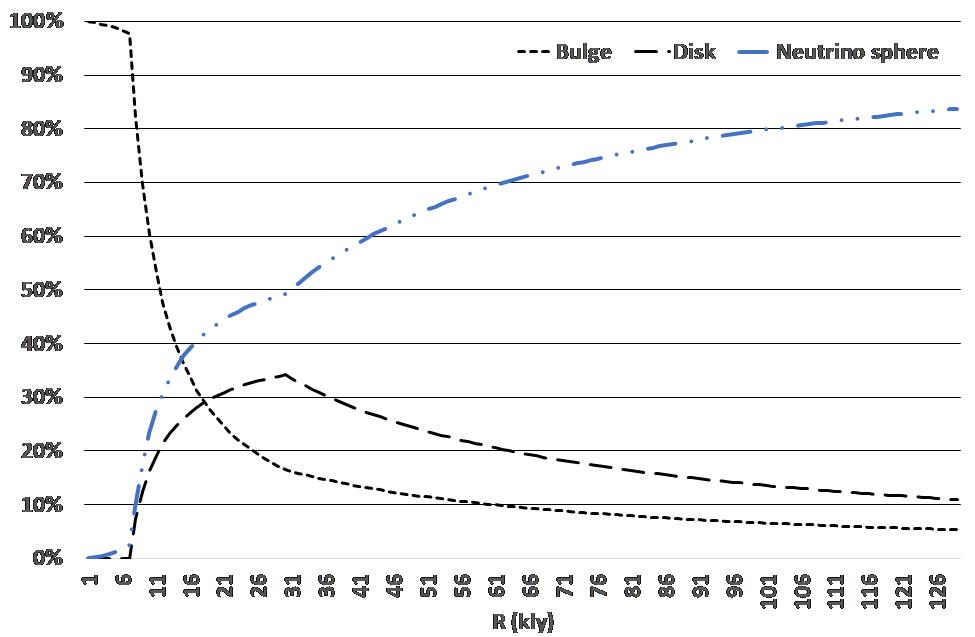}
		\caption{Est. constituents' mass profiles.}	
	\end{minipage}
	\begin{minipage}[b]{.5\textwidth}
		\includegraphics[width=0.99\linewidth]{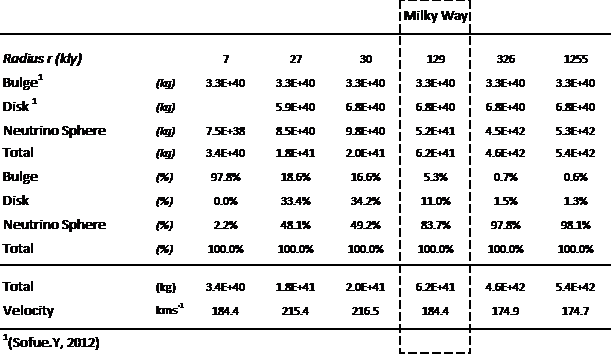}
		\caption{Est. mass-velocity profiles.}	
	\end{minipage}
\end{figure}

\begin{figure}[h]
	\centering
	\begin{minipage}[b]{.99\textwidth}
		\includegraphics[width=0.99\linewidth]{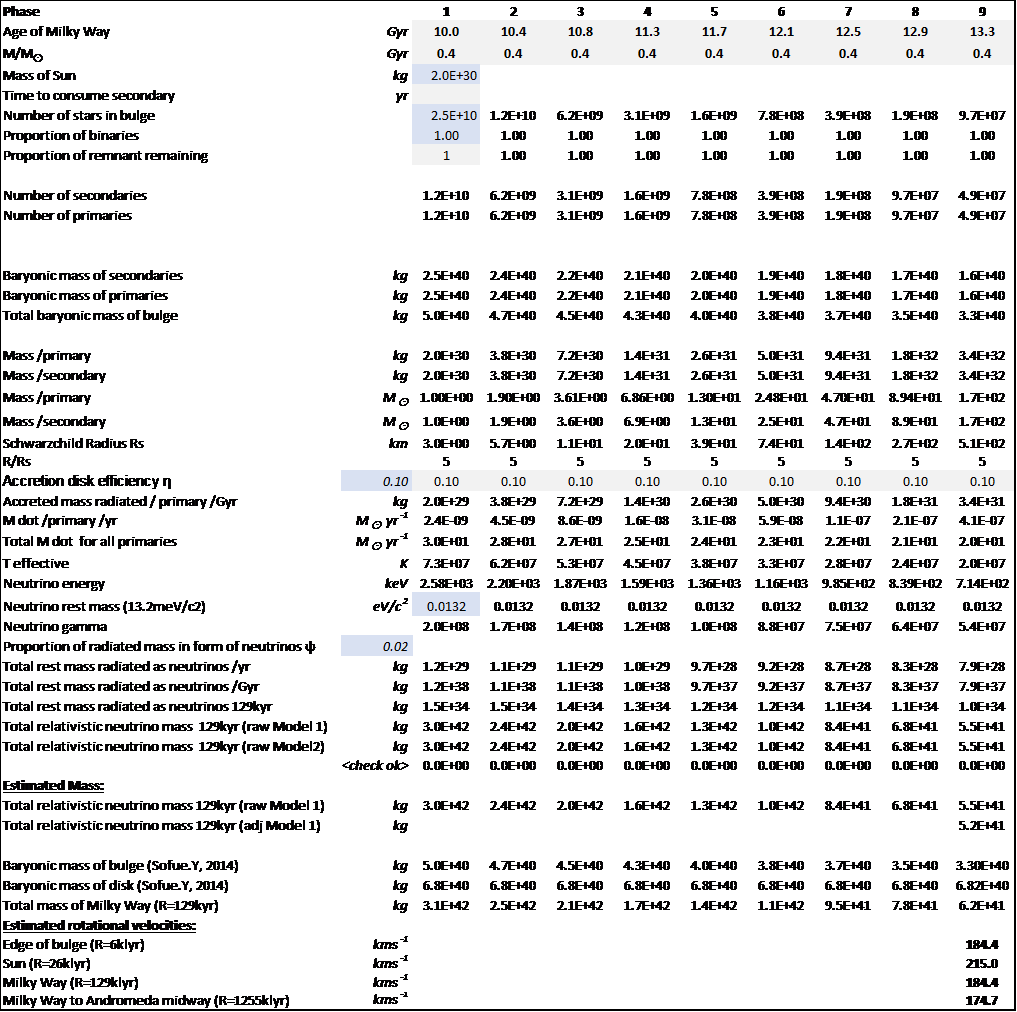}
		\caption{Model: summary parameters \& results.}	
	\end{minipage}
\end{figure}

\noindent \textbf{Discussion}\\

\noindent The estimated rotational velocity curve (Fig.3) appears to be consistent with the observed velocity curve (Fig.1) over a distance of~$\sim70kly$. For distances beyond this, the dispersion of observed velocities for a given distance appears to increase significantly (Fig.2), thereby affecting the significance of the observations - after initially appearing to decline, velocities subsequently increase to a level of$\sim150kms^{-1}$ at a distance of $385kpc$ compared to$\sim175kms^{-1}$ for the model. One of the possible causes of this may be that the model does not consider other gravitational potentials not related to the Milky Way. Given that this distance corresponds to approximately the mid-point between the Milky Way and the similarly sized M31, it may be close to a Lagrange point in the overall gravitational field, which may have an impact on the rotation curve when viewed relative to the Milky Way.
\begin{wrapfigure}{r}{2.65in}
	\begin{minipage}{.38\textwidth}
		\centering	
		\includegraphics[scale=0.8]{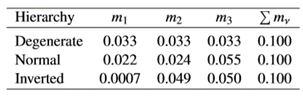}
		\caption{Neutrino masses (eV).\cite{palanque2015neutrino}}	
		\end{minipage}\\
\end{wrapfigure}
\noindent The reasonable fit of the estimated velocity curve with the observed velocities (Fig.4) is also consistent with electron neutrinos in their normal hierarchical state of having a rest mass of of$\sim13meV/c^2$, which does not appear inconsistent with~\cite{palanque2015neutrino}, who suggest the rest masses of the three neutrino species can follow a normal hierarchy with two light states and a heavier one, in which case the minimum total rest mass would be $\sim$ $\sum m_\nu=0.06 eV$. Their suggested masses of the three species by hierarchy for $\sum m_\nu=0.10eV$ are shown in Fig.10, for which $\sum m_\nu=0.06eV$ would suggest that an electron neutrino in the normal hierarchy may have a rest mass of $\sim 0.022x0.06/0.10eV$ or $\sim13meV/c^2$. This value was utilised in the model and generated a "dark matter" neutrino sphere mass corresponding to $\mathrm{\sim }$84\% of total mass for the Milky Way (at $R=129kly$), which appears to be consistent with the observed rotational velocity curve~\cite{sofue2012grand}.\\

\noindent \textbf
{Conclusion}\\

\noindent The gravitational effect of stellar neutrinos produced during a typical stellar main sequence nucleosynthetic process appears to be insignificant compared to stellar baryonic mass due to a combination of relatively low $\dot{M}$ and $\eta$ associated with this process. Conversely, however, the model herein suggests a mechanism whereby concentration of matter associated with mass-radiating accretion disks, such as potentially at the centre of the Milky Way, could provide the relatively higher $\dot{M}$ and $\eta$ required to generate relativistic neutrinos with sufficient flux and energies to have a gravitational effect significantly higher than that due to baryonic mass and that such mass could be sufficient to give a reasonable account of the observed observed rotational velocity curves (Fig.4) and the estimated proportion of dark matter to baryonic matter$\sim84\%$, consistent with current physical laws of gravitation. The model additionally implies that the observed rotational dynamics of the Milky Way is consistent with a rest mass of the electron neutrino in the normal hierarchy of$\sim13meV/c^2$, consistent with current understanding\cite{palanque2015neutrino}.\\

\noindent As, ``the rotational velocity curve of the Milky Way is not atypical of other spiral galaxies''~\cite{carroll2017introduction}, this may suggest that such a mechanism may also be a systematic galactic phenomenon. However, the model implies that until such galaxies reach an apparent age of $\sim10Gyr$ required for Phase I of the accretion-burning neutrino sphere to commence, galaxies should, everything else being equal, exhibit a relatively insignificant proportion of neutrino sphere mass relative to baryonic mass, when compared to their apparently ``older'' counterparts. This is consistent with the observation that the proportion of "dark matter" mass to baryonic mass is insignificant for galaxies of relatively high red shifts z$\sim0.6-2.5$~\cite{genzel2017strongly}, corresponding to an apparent distance of at least $\sim5Glyr$ away, assuming a Hubble "constant" of $76kms^{-1}$, compared to an apparent distance of $\sim4Glyr$ away implied my the model.\\

\noindent An additional implication of the model is that the radiative neutrino spheres of galaxies less than$\sim3.5Glyr$ away would act as mass-radiating beacons that would be expected to create mass-interference patterns with spheres radiated by other galaxies to potentially create an overall stationary superposition of radiated "mass" to which baryonic matter, such as gas and dust in the intergalactic medium, may clump to, not dissimilar to observed dark matter filaments~\cite{riordan1991shadows}.\\

\noindent Finally, the model implies two relatively different apparent mass distribution functions for the observable universe corresponding to $t<10Gyr$ and $t>10Gyr$ respectively. The denser mass function of the latter relating to observable galaxies that are apparently relatively closer by, suggests a relatively greater enclosed mass density of nearby observable galaxies compared to those more "baryonic" galaxies observed to be relatively farther afield, thereby potentially creating an apparent observed acceleration effect on galaxies with distance that may be associated with "dark energy".\\ 

\noindent Further research would need to be undertaken to confirm and refine the mechanism herein, to evaluate its consistency with other observable astrophysical galactic and extragalactic phenomena and to explore the suggested potential implication on "dark energy".

\bibliographystyle{plain}
\bibliography{references.bib}

\end{document}